\documentclass[%
reprint,
amsmath,amssymb,
aps,
pra,
longbibliography,
]{revtex4-1}

\usepackage{graphicx}
\usepackage{dcolumn}
\usepackage{bm}
\usepackage{natbib}
 \usepackage{ booktabs}

\begin{document}

\title{Mitotic waves in an import-diffusion model with multiple nuclei in a shared cytoplasm}

\author{Felix E. Nolet$^{1}$}
\author{Lendert Gelens$^{1}$}

\affiliation{%
\textbf{\\}
{$^1$Laboratory of Dynamics in Biological systems\text{,} Department of Cellular and Molecular Medicine, Faculty of Medicine, KU Leuven (Belgium)}
}%

\date{\today}

\begin{abstract}
Nuclei import and export proteins, including cell cycle regulators. These import-export processes are modulated periodically by the cell cycle, for example due to the periodic assembly and breakdown of the nuclear envelope. As such, replicated DNA can be segregated between the two daughter cells and the proteins that were localized in the nucleus are free to diffuse throughout the cytoplasm. Here, we study a mathematical import-diffusion model to show how proteins, i.e. cell cycle regulators, could be redistributed in the cytoplasm by nuclei that periodically toggle between interphase and mitosis. We show that when the cell cycle period depends on the local concentration of regulators, the model exhibits mitotic waves. We discuss how the velocity and spatial origin of these mitotic waves depend on the different model parameters. This work is motivated by recent \textit{in vitro} experiments reporting on mitotic waves in cycling cell-free extracts made with \textit{Xenopus laevis} frog eggs, where multiple nuclei share the same cytoplasm. Such experiments have shown that nuclei act as pacemakers for the cell cycle and thus play an important role in collectively defining the spatial origin of mitotic waves. 
\end{abstract}




\maketitle

\section{Introduction}
In eukaryotic cells, progression through the cell cycle depends on the correct termination of previous events~\cite{Murray1989science}. For example, a cell has to grow sufficiently before DNA replication can start, and all DNA has to be replicated before mitosis starts. However, these cell division cycles are different in the early embryos of insects, amphibians, and fish, which lay their eggs externally. After fertilization, these organisms go from a single, large cell to several thousands of somatic-sized cells as quickly as possible by carrying out multiple rounds of rapid cleavages~\cite{FoeAlberts1983,Olivier2010,Farrell2014,Anderson2017}. The cell cycle in these early embryos resembles a clock driven by an autonomous biochemical oscillator~\cite{Murray1989nature}. In what follows, we will focus our attention on cell cycle oscillations in the early embryo of the frog \textit{Xenopus laevis}, which is about $1$ mm in diameter, and its oscillation period is around 25 min. 

The main driver of the cell cycle is the kinase Cdk1 (cyclin-dependent kinase 1). When Cdk1 is bound to cyclin B, it can phosphorylate many substrates, leading to mitosis. The activity of the cyclin B-Cdk1 complex is further regulated by phosphorylation. Positive feedback loops involving phosphatases, such as Cdc25, and kinases, such as Wee1, have been shown to lead to bistability~\cite{Pomerening2003,Moore2003}. This bistability means that for some (constant) concentrations of cyclin B, there exist two stable states of cyclin B-Cdk1 activity. One of the substrates of Cdk1 is the APC/C (Anaphase-Promoting Complex/Cyclosome), an E3 ubiquitin ligase, which has recently been shown to also respond in a bistable manner to Cdk1~\cite{Mochida2016,Kamenz2021}. The activation of APC/C leads to the destruction of cyclin B, and this negative feedback loop (cyclin B-Cdk1 $\rightarrow$ APC/C $\dashv$ cyclin B-Cdk1) is at the heart of biochemical oscillations in cyclin B-Cdk1 activity, which was already mathematically described in the 90s~\cite{Goldbeter1991,Tyson1991,Novak1993}.

However, describing cell cycle oscillations only in time does not give the complete picture, since the concentrations or activities of the proteins involved might also vary in space. For example, the concentration or activity of cell cycle regulators can differ between the nucleus and the cytoplasm. Cyclin B-Cdk1 accumulates rapidly in the nucleus before the nuclear envelope breaks down~\cite{Santos2012,Pines2010}. Cdc25 is also translocated to the nucleus at the start of M phase~\cite{Toyoshima-Morimoto2002}, but Wee1 is mostly present in the nucleus during S phase~\cite{Heald1993}. In other words, the concentration and spatial distribution of cell cycle regulators depend on the cell cycle itself. This can be incorporated into mathematical models in different ways. One way to include spatial dynamics is to add one or more spatial dimensions and diffusion to the existing temporal models. As such, one moves from \textit{ordinary} differential equations to \textit{partial} differential equations. Differences in the dynamics at certain spatial locations could then be introduced through the parameters in such a model. One could also introduce different cellular compartments and the interactions between these compartments, potentially assuming that proteins are well-mixed within each compartment. It has been shown that such compartmentalized models can exhibit bistability -- or multistability in general -- when the model in the original setting does not~\cite{Harrington2013}. In the context of the cell cycle, the presence of two compartments (e.g. nucleus and cytoplasm) can lead to bistable switches that dynamically change throughout the cell cycle, making cell cycle oscillations more robust~\cite{Rombouts2021,Rombouts2021Arxiv}. Two-compartment models have also allowed to explain pattern formation processes in different biological systems~\cite{Brauns2020}. 

\begin{figure*}[h!]
	\begin{center}
		\includegraphics[width=\textwidth]{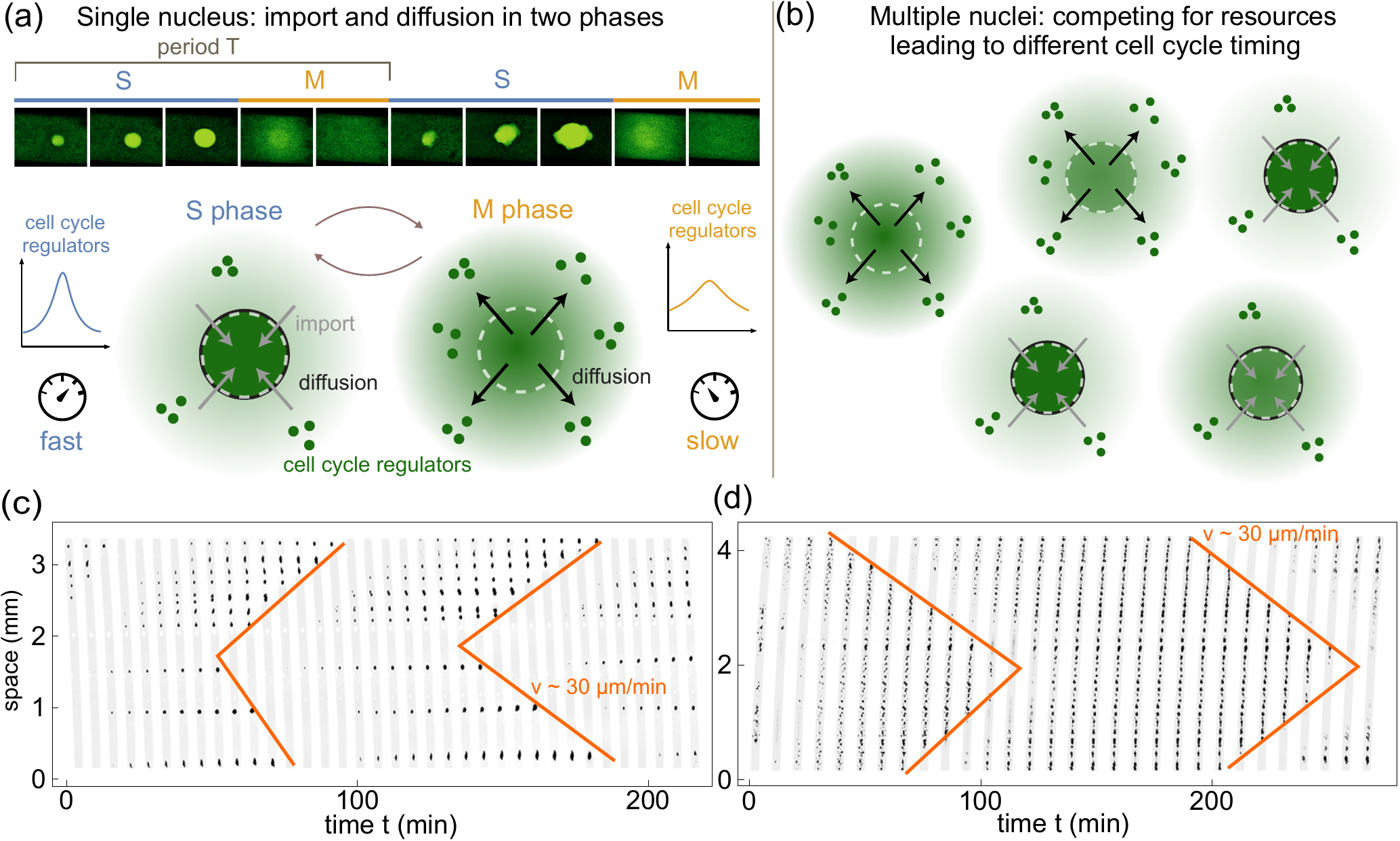}
	\end{center}
	\caption{
		 (a) Snapshots of measured fluorescence in a part of a tube (image dimensions: $290\times230$ $\mu$m), zoomed in at one nucleus going through S and M phase periodically. Below a sketch of the two phases of the nucleus: (i) S phase with both import and diffusion, a high concentration of cell cycle regulators and thus a fast cell cycle, and (ii) M phase with only diffusion, a lower concentration in the nucleus and a slower cell cycle. (b) Sketch of multiple nuclei competing for cell cycle regulators. Nuclei can have different cycle speeds and thus be in different phases of the cycle. (c) Images of multiple nuclei in full Teflon tube, with time in the horizontal direction. Mitotic waves (disappearance of nuclei, orange lines) start internally. (d) Similar as (c) but the mitotic waves originate at the boundary of the tube.}
	\label{fig1}
\end{figure*}

Cycling cell-free extracts made from \textit{Xenopus laevis} frog eggs provide a convenient experimental system to controllably test how cell cycle oscillations are coordinated spatially. By adding nuclear material (sperm chromatin) to the extract, many nuclei spontaneously self-organize within the shared cytoplasm. One way to visualize nuclei is to use a fluorescent reporter, such as GFP-NLS (green fluorescent protein with a nuclear localization signal). These extracts continue to biochemically oscillate in Cdk1 activity. As a result, nuclei are visible during S phase, but disappear in M phase when the nuclear envelope breaks down and the GFP-NLS diffuses away. This can be seen in Fig.~\ref{fig1}(a), where the fluorescence around one nucleus in the extract is plotted at different times. 
From the measured fluorescence in these experiments, we estimate a diffusion coefficient $D$ of GFP-NLS in the cytoplasm to be around $600 - 1200$ $\mu$m$^2$/min. This estimate is obtained by approximating the intensity at the end of S phase with a Gaussian function and fitting this to the solution of the diffusion equation during M phase. Since the cyclin B-Cdk1 complex has a larger molecular weight, the corresponding diffusion coefficient is likely slightly smaller. More precisely, the diffusion coefficients estimated based on the molecular weights differ approximately with a factor of 1.15, which is smaller than the uncertainty in our estimate of $D$ (the estimated range for $D$ of cyclin B-Cdk1 is then around $500 - 1000$ $\mu$m$^2$). The effective diffusion coefficient could also differ due to binding and interaction with other molecules in the cell, and due to changes in its shape (e.g. folding and unfolding of proteins)~\cite{Guo2014,Korvasova2015}. Therefore this should be seen as a rough estimate. 

When visualizing the dynamics of multiple nuclei in the extract over time, waves of mitosis have been observed~\cite{Chang2013}. An example is shown in Fig.~\ref{fig1}(c) where the mitotic entry (observed via the disappearance of nuclei) is coordinated by a wave propagating at a constant velocity of about 30 $\mu$m/min. Traveling waves are commonly observed in a wealth of biological processes, ranging from action potentials in neurons to chemical reactions such as the Belousov - Zhabotinsky reaction~\cite{HodgkinHuxley,Zaikin1973,Winfree1987,Tyson1988,Gelens2014,Kruse2017,Deneke_review_2018}. Such waves can transmit information over large distances in a fast and robust way. This is especially relevant in large developing eggs such as those of \textit{Xenopus laevis}, because they are too large to be synchronized by diffusion alone~\cite{Chang2013,Gelens2014}. Mitotic waves have also been observed \textit{in vivo} in the syncytium of the \emph{Drosophila melanogaster} embryo ~\cite{Deneke2016,Vergassola2018}. Recent work has shown that nuclei actively play a role as pacemakers for the cell cycle, and multiple nuclei collectively determine the spatial origin of mitotic waves ~\cite{Afanzar2020,Nolet2020}. While in Fig.~\ref{fig1}(c), a mitotic wave was triggered in the center of the domain, it has also been regularly observed to originate from the system boundary (see Fig.~\ref{fig1}(d)). It has been hypothesized that nuclei serve as pacemakers by concentrating cell cycle regulators, which in its turn increases the local frequency of cell cycle oscillations~\cite{Nolet2020}.

In this paper, we study a mathematical import-diffusion model for the concentration of an unspecified cell cycle regulator, where nuclear import is periodically modulated by the cell cycle oscillator. The model assumes that a nucleus locally increases the concentration during a part of the cell cycle (S phase). In mitosis (M phase), the nucleus disappears, and all regulators are free to diffuse throughout the cytoplasm (Fig.~\ref{fig1}(a)). We introduced a similar model in \cite{Nolet2020}, but generalize it here and study its dynamics in more detail. We alter the number of nuclei, ranging from a single nucleus to multiple nuclei (Fig.~\ref{fig1}(b)), as motivated by the experiments in \textit{Xenopus} extracts (see Fig.~\ref{fig1}(c)-(d)) and the early \emph{Drosophila} embryo. By assuming that the cell cycle period depends on the local concentration of cell cycle regulators, the model shows mitotic waves. Moreover, the origin of the waves can be controlled by the nuclear positioning. Although we focus on cell cycle oscillations, the model itself is generic. It essentially describes any continuous system where a concentration locally increases and then diffuses away periodically at one or more positions, where that concentration is possibly coupled to the frequency of the oscillation. 

In the next section, we start with the definition of the model. After that, we study the situation where the cell cycle period is constant everywhere in space. Finally, we study the effects of a changing cell cycle period by assuming that it depends on the local concentration of cell cycle regulators.

\section{Definition of the model}
\label{sec:modeldef}
The model describes the concentration of proteins (cell cycle regulators) in the presence of diffusion and nuclear import. The concentration is denoted by $C(x,t)$ and depends on time ($t$) and space ($x$). This is a generic concentration of any cell cycle regulating protein, with as the only assumption that this protein is spatially localized during the cell cycle and that its local concentration in its turn could change the period of the cell cycle. Examples of such proteins are Cdc25 and Wee1, which affect the cell cycle via positive feedback with the Cdk1-Cyclin B complex. The equation for the change of $C(x,t)$ is given by
\begin{equation}
\label{eq:model1}
\frac{\partial C(x,t)}{\partial t}=D\nabla^2 C(x,t)+\nabla\cdot(C(x,t)\nabla V(x,t)),
\end{equation}
with the dot denoting the inner product. The change of the concentration consists of two terms: diffusion (with diffusion coefficient $D$) and attraction via the potential function $V(x,t)$. In general, multiple nuclei are present ($N\geq 1$) and $V(x,t)$ describes the attraction of proteins to the nuclear positions $\{\xi_i\}_{i=1,...,N}$, mimicking nuclear import without the need of explicitly defining boundaries of the nuclei. Every nucleus defines such a potential $V_i(x,t)$ and together they form the full potential function
\begin{equation}
V(x,t)=\sum_{i=1}^{N}V_i(x,t)
\end{equation}
via superposition. For every separate nucleus the potential is written as
\begin{equation}
V_i(x,t)=F_i(t)G_i(x),
\end{equation}
i.e. it can be separated in a time-dependent ($F_i$) and a space-dependent ($G_i$) function. The function $G_i(x)$ determines the attraction at this nucleus. Whereas the exact shape can be chosen freely, a rather standard function to describe the attraction is a Gaussian exponential function,
\begin{equation}
G_i(x)=-\epsilon_i e^{-\frac{(x-\xi_i)^2}{\sigma_i^2}}.
\end{equation}
This function has parameters $\epsilon_i$ and $\sigma_i$, altering the depth and width of the potential function, respectively. Biologically this means that $\epsilon_i$ is a measure for the strength of the attraction at (or equivalently, import into) a nucleus, and $\sigma_i$ a measure for the attraction width (i.e. from how far away can a nucleus import proteins). The function $F_i(t)$ ensures that the attraction at the nuclei is only present during a fraction ($0<\alpha<1$) of the cell cycle, i.e. during S phase. 
It is essentially periodically turned on and off for each nucleus. It is defined by
\begin{equation}
F_i(t)=\left\{\begin{array}{ll}
1&\text{if } \phi_i(t) \text{ mod } T_0 <\alpha T_0\\
0&\text{if } \phi_i(t) \text{ mod } T_0 \geq \alpha T_0
\end{array}\right.
\end{equation}
where $T_0$ is the reference value for the cell cycle period and $\phi_i(t)$ denotes the so-called \emph{phase} of each nucleus, defined by
\begin{equation}
\frac{d\phi_i}{dt}=v_i(t),
\end{equation}
changing through time with a speed $v_i(t)$. At $t=0$, we define the phase to be $\phi_i(0)=0$ for all nuclei. However, the speed $v_i(t)$ can change over time as it is coupled to the local concentration $C(x,t)$ as follows:
\begin{equation}
\label{eq:vt}
v_i(t)=1+\eta(C(\xi_i,t)-C_0)
\end{equation}
where $C_0\geq0$ a reference concentration and $\eta$ a constant that determines how strongly the phase speed is coupled to the concentration at the  location of the i$^{th}$ nucleus $\xi_i$. The underlying biological assumption is that nuclei that import (attract) more cell cycle regulators have a shorter cell cycle length than other nuclei. Note that for $\eta=0$, we have $v_i=1$ for all nuclei and thus obtain $\phi_i(t)=t$. This implies that the model becomes uniform and all nuclei are in S phase or M phase at the same time, with cell cycle period $T=T_0$.

In the model, we assume zero-flux boundary conditions, i.e. proteins cannot move outside the domain. We can rewrite the model equation as a flux equation,
\begin{equation}
\frac{\partial C(x,t)}{\partial t}+\nabla\cdot\vec{J}(x,t)=0,
\end{equation} 
where $\vec{J}(x,t)$ denotes the flux of $C(x,t)$. For our model we find the flux to be
\begin{equation}
\vec{J}(x,t)=-(D\nabla C(x,t)+C(x,t)\nabla V(x,t)).
\end{equation}
The boundary condition can then be written as $\vec{J}(x,t)\cdot\vec{n}=0$ at the boundary, with $\vec{n}$ denoting the normal vector at the boundary. For numerical details, we refer to Appendix~\ref{ap:numerics}.

\begin{table}[b!]
	\begin{center}
		\begin{tabular}{ | c | p{42mm} | c | c | }
			\hline			
			par. & explanation & value & unit \\ \hline\hline
			$D$ & diffusion coefficient & 600 &  $\mu$m$^2$/min	\\ \hline 
			$T_0$ & reference cell cycle period & 40 & min  \\ \hline
			$\alpha$ & fraction of S phase & 0.7 &	\\ \hline 
			$\epsilon$ & attraction strength & 300 &  $\mu$m$^2$/min \\ \hline
			$\sigma$ & (measure for) attraction width & 100 & $\mu$m \\ \hline
			$\eta$ & coupling coefficient cell cycle to concentration & 0 -- 1 & min$^{-1}$ \\ \hline
			$C_0$ & reference concentration & 1 & \\ 
			\hline  
		\end{tabular}
	\end{center}
	\caption{
		Standard parameter values and their units.}
	\label{tab1}
\end{table}

To initialize the model, we take $C(x,0)=1$ for all $x$ and define the positions of all nuclei ($\{\xi_i\}_{i=1,...,N}$) which remain fixed throughout the simulation. Strictly speaking, the equation itself is non-dimensional, i.e. there are no units attached to the variables and parameters. However, inspired by the biological context (Fig.~\ref{fig1}), we think of time $t$ in minutes, and space $x$ in micrometers. In this way, we can compare the output of the model with experimental results and, vice versa, choose the value of certain parameters. For example, the cell cycle period of the early \emph{Xenopus laevis} embryo is in the order of 20--60 minutes (it is typically slower in extracts than \textit{in vivo}), and from experiments (Fig.~\ref{fig1}) we estimated the diffusion coefficient of Cdk1 to be in the range of 500--1000 $\mu$m$^2$/min. For \textit{Xenopus laevis}, the standard values of the model parameters and their units are given in Table~\ref{tab1}. These parameter values are estimated from experimental data such as in Fig.~\ref{fig1} and have been used in previous experimental and theoretical research~\cite{Nolet2020}.

In the next two sections we discuss the results of the model for a constant cell cycle period ($\eta=0$) and a concentration dependent cell cycle period ($\eta>0$), respectively.

\begin{figure*}[h!]
	\begin{center}
		\includegraphics[width=\textwidth]{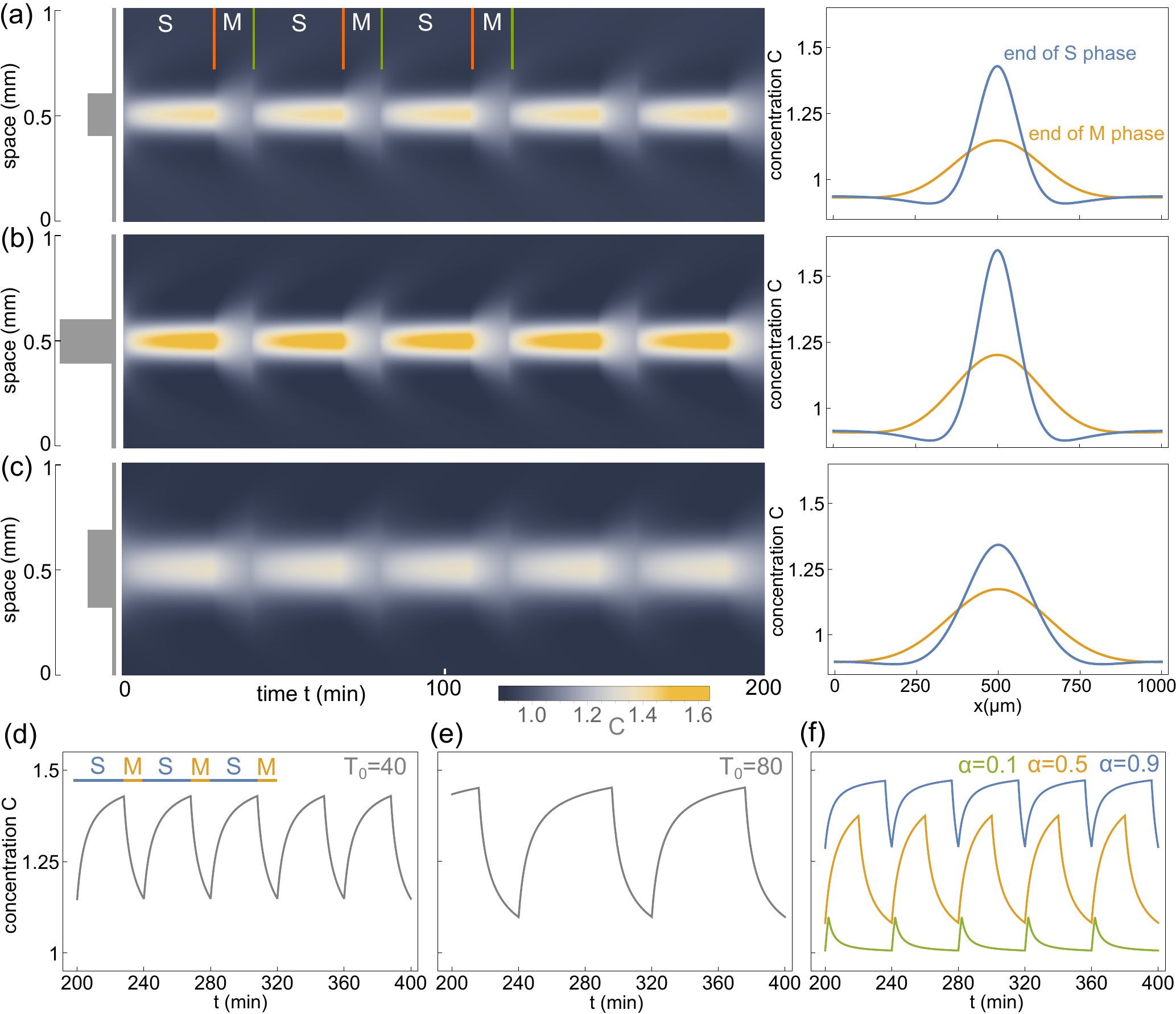}
	\end{center}
	\caption{
		Single nucleus at $\xi=500$ $\mu$m in a domain of size $L=1000$ $\mu$m, for different values of nuclear parameters $\epsilon$ and $\sigma$. Cell cycle parameters are fixed at $\alpha=0.7$ and $T_0=40$ min. (a) Left: a sketch of the nucleus (with standard nuclear parameters $\epsilon=300$ $\mu$m$^2$/min and $\sigma=100$ $\mu$m) and the corresponding concentration in space and time for five cycles (dimensions: $1000$ $\mu$m $\times$ $200$ min). Right: a graph of the concentration in space at the end of S phase (orange) and M phase (green). (b) Similar, for increased $\epsilon=400$ $\mu$m$^2$/min. (c) Similar, for increased $\sigma=160$ $\mu$m. (d) Time series of $C$ at the center ($x=\xi=500$ $\mu$m), corresponding to (a). (e) Similar graph for an increased period $T_0=80$. (f) Similar graph for different values of $\alpha$.}
	\label{fig2}
\end{figure*}

\section{Model with constant cell cycle period}
In this section, we study the model with a constant cell cycle period, i.e. the case $\eta=0$. This means that the transition between S phase and M phase occurs simultaneously for all nuclei everywhere in space, independent from the local concentration. Note that this requires the fact that we also start the cell cycle at the same time everywhere in the system. We start our analysis with the case for a single nucleus ($N=1$), after which we discuss the results for multiple nuclei ($N>1$).

\subsection{Single nucleus}
We consider an interval $[0,L]$ of length $L$ in which we place a nucleus in the middle, at $\xi=L/2$. For now, we fix the domain size to $L=1000$ $\mu$m and choose the cell cycle period to be $T_0=40$ min. We assume that S phase, where regulators can be imported into the nucleus, accounts for 70\% of the cell cycle, so $\alpha=0.7$. 
Furthermore, the diffusion constant is taken to be $D=600$ $\mu$m$^2$/min. 
Fig.~\ref{fig2}(a)-(c) shows the simulation results for different values of the ``import strength" $\epsilon$ and the ``import range" $\sigma$ of the nucleus. In all cases, there is an increase of concentration in S phase due to the attraction, and a decrease in M phase due to diffusion. The concentration profiles at the end of S and M phase are shown on the right. When increasing the import strength ($\epsilon$), the concentration profile at the end of S phase increases due to the stronger attraction at the nucleus (see Fig.~\ref{fig2}(a) vs. Fig.~\ref{fig2}(b)). When the range of attraction of the nucleus ($\sigma$) is increased, the concentration profile widens and the concentration difference between S phase and M phase becomes smaller (see Fig.~\ref{fig2}(a) vs. Fig.~\ref{fig2}(c)).

Instead of changing the properties of the nucleus (via $\epsilon$ and $\sigma$), we also changed the cell cycle oscillation parameters, such as the oscillation period $T_0$ and the relative duration of S phase to the total period ($\alpha$). Fig.~\ref{fig2}(d) shows how the concentration at the nucleus changes in time for the same simulation as in Fig.~\ref{fig2}(a). The concentration peaks at the end of S phase, while it is minimal at the end of M phase. 
When doubling the oscillation period to $T_0=80$ min, the amplitude of the oscillation increases (Fig.~\ref{fig2}(e)). Indeed, since both S phase and M phase are doubled in time, the attraction in S phase lasts longer leading to a higher concentration at the end of S phase. Similarly, the concentration at the end of M phase is lower since the period during which there is only diffusion has doubled as well. Fig.~\ref{fig2}(f) illustrates the effect of changing $\alpha$, while keeping the period $T_0$ fixed. When increasing the fraction of S phase to $\alpha=0.9$ (blue), we see that the entire curve shifts upwards and the amplitude decreases. The period during which the proteins can diffuse in the absence of nuclear attraction in M phase is now too short for the concentration to return to lower levels. When decreasing the fraction to $\alpha=0.5$ (orange), the curve shifts downwards for similar reasons. For $\alpha=0.1$ (green), the curve shifts down even further and the amplitude decreases as the period of nuclear attraction is much shorter. 

\begin{figure*}[h!]
	\begin{center}
		\includegraphics[width=\textwidth]{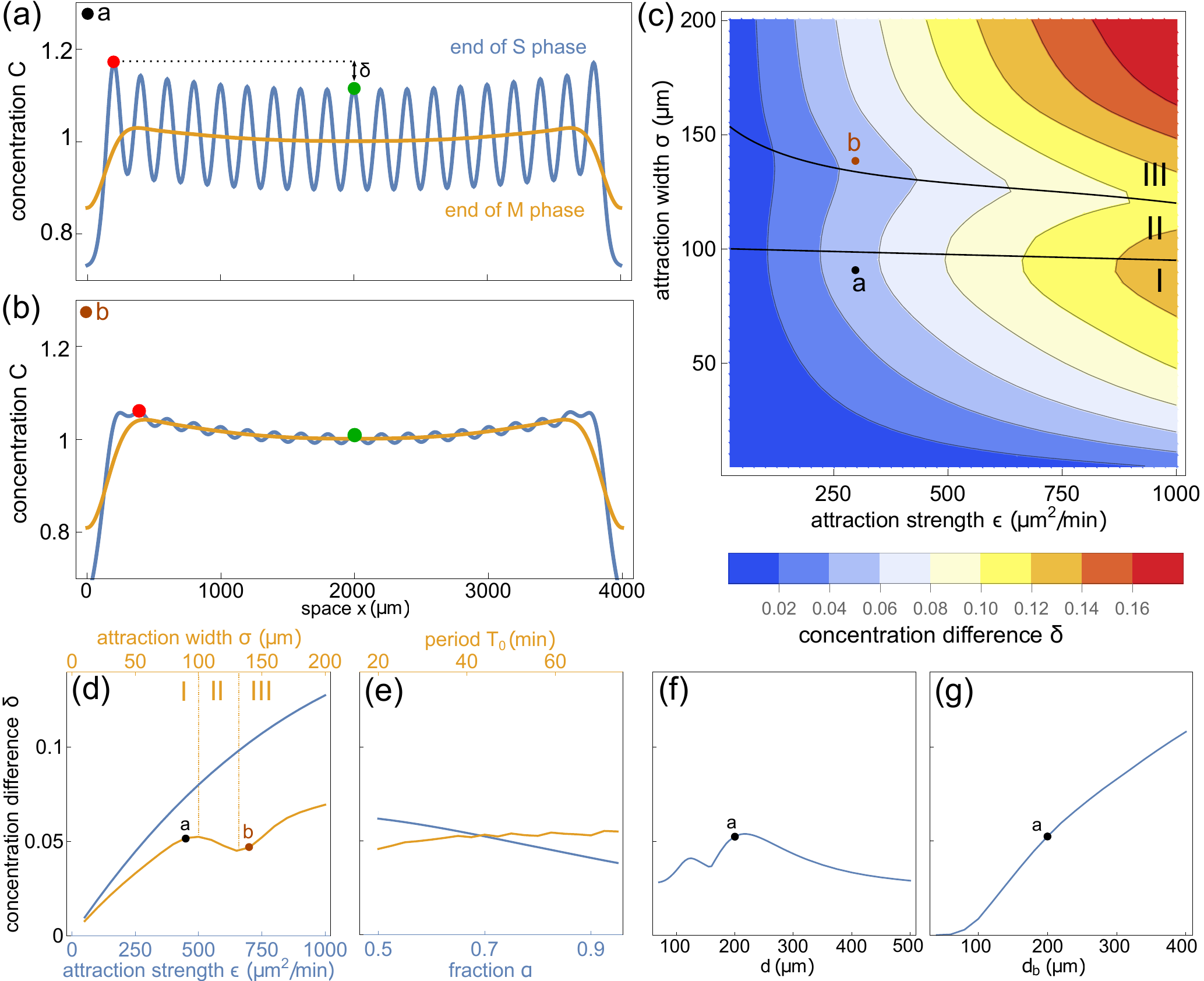}
	\end{center}
	\caption{
		Simulations with multiple (19) equidistant nuclei. If not stated otherwise, parameter values are as in Table~\ref{tab1}. (a) Concentration at the end of S phase (blue) and the end of M phase (orange) as function of space for $\sigma=90$ $\mu$m. The red dot denotes the maximum concentration, the green dot the concentration in the center, at the end of S phase. The relative concentration difference is denoted by $\delta$. (b) Similar graph for $\sigma=140$ $\mu$m. (c) Heatmap of the relative concentration difference $\delta$ for varying $\epsilon$ and $\sigma$. The black lines divide the parameter space into three domains, where the maximum concentration is located (I) at the first nucleus (200 $\mu$m), (II) right from the first nucleus, but still being the closest (200-300 $\mu$m) or (III) closer to the second nucleus (>300 $\mu$m). The black and brown dot (a,b) correspond to the graphs in (a) and (b). (d) Relative concentration difference as function of nuclear parameters $\epsilon$ and $\sigma$. (e) Relative concentration difference as function of cell cycle parameters $\alpha$ and $T_0$. (f) Relative concentration difference as function of internuclear distance $d$. (g) Relative concentration difference as function of $d_b$, the distance of the outer nuclei to the boundary while keeping the internuclear distance fixed at $d=200$ $\mu$m.}
	\label{fig3}
\end{figure*}

\subsection{Multiple nuclei}
Now that we have a better understanding of the behavior of a single nucleus in this model, we study the situation with multiple nuclei. However, since the potential function $V(x,t)$ is then the superposition of potentials of individual nuclei $V_i(x,t)$, we have to pay attention to the positioning of the nuclei. More precisely, there is an interplay between the distance between nuclei and the attraction width $\sigma$. For a large internuclear distance (or small $\sigma$) the individual potentials $V_i(x,t)$ barely overlap and all nuclei act independently (as described in the previous section). However, for a small internuclear distance (or large $\sigma$), the potentials overlap so much that the resulting potential resembles one large nucleus instead of multiple interacting nuclei. Therefore, we choose the distance and width such that we are in the interesting regime where the potentials $V_i(x,t)$ only significantly overlap with the immediate neighbors. In this case, we study how the competition between multiple nuclei affects the spatial concentration profiles.

We set up the system with $N$ equidistantly distributed nuclei, all separated by a distance $d$:
\begin{equation}
\{\xi_i\}_{i=1,...,N}=\{id\}_{i=1,...,N}=\{d,2d,...,Nd\}.
\end{equation} 
Moreover, we choose the distance of the outermost nuclei to the boundary of the domain also to be equal to $d$. This yields a simple relation between the distance $d$ and domain size $L$:
\begin{equation}
d=\frac{L}{N+1}.
\end{equation}

Similar to the experimental measurements using cell-free \textit{Xenopus laevis} extracts shown in Fig.~\ref{fig1}, we set $L=4000$ $\mu$m with 19 nuclei, thus $d=200$ $\mu$m. Fig.~\ref{fig3}(a) shows the resulting dynamics. At the end of S phase, the concentration peaks at the locations $\xi_i$ of the nuclei. However, the nuclei towards the boundary have a higher concentration than the ones in the center of the domain. These nuclei are different in the sense that they lack a second neighbor and therefore less competition for resources (less overlap in potential $V$). Having higher concentrations towards the boundaries is also observed in the average GFP intensity in experiments with \textit{Xenopus laevis} extracts~\cite{Nolet2020}. We define $\delta$ as the difference between the maximum concentration (red dot) and the concentration in the center (green dot), relative to the latter. For $\delta=0$ the maximum is attained at the center, and for $\delta>0$ it is towards the boundaries. When increasing the attraction range $\sigma$, the concentration peaks at the nuclei are much less pronounced and the location of maximal concentration moves more towards the second nucleus (Fig.~\ref{fig3}(b)). Figure~\ref{fig3}(c)-(d) shows the relative concentration difference $\delta$ as function of nuclear parameters, i.e. the attraction strength $\epsilon$ and attraction range $\sigma$. The concentration difference increases with the attraction strength $\epsilon$, since the nuclei are able to import more regulators during S phase (including those at the boundary). The parameter space is divided into three domains for the location of the concentration maximum at the end of S phase, which we denote by $x_{\text{max}}$. In region I, the maximum is located at the first nucleus ($x_{\text{max}}=200$ $\mu$m). In region II, the maximum is to the right of first nucleus, but still closest to that nucleus, i.e. $200<x_{\text{max}}<300$ $\mu$m. In region III, the maximum is closer to the second nucleus ($x_{\text{max}}>300$ $\mu$m). This phenomenon, as described also in Fig.~\ref{fig3}(b), mainly depends on the attraction width $\sigma$. The points labeled a and b, in regions I and III, correspond to Fig.~\ref{fig3}(a) and Fig.~\ref{fig3}(b). Fig.~\ref{fig3}(e) similarly shows how the relative concentration difference $\delta$ changes, but as function of the cell cycle parameters $\alpha$ (S phase fraction) and $T_0$ (cell cycle period). The relative concentration difference $\delta$ decreases with $\alpha$, but does not have a very clear dependence on the period $T_0$. Overall, the dependence on the nuclear parameters is more pronounced. 

Next, we explored the influence of the internuclear distance $d$, keeping the same number of nuclei. The concentration difference $\delta$ varies with $d$ and is maximal around $d=220$ $\mu$m (Fig.~\ref{fig3}(f)). For larger values of $d$, the concentration difference decreases. Indeed, when nuclei are further apart, there is less competition for the common pool of proteins, and they approach similar dynamics as individual nuclei (see previous section). For smaller internuclear distances $d$, $\delta$ generally decreases, except for a range of distances $d$ where the maximum concentration at the end of S phase is not located at the outermost nucleus, but closer to the second (region II). Note that this behavior is similar as for changing $\sigma$, while keeping the distance $d$ fixed. 

Lastly, we wondered how the distance between the outermost nuclei and the boundary, denoted by $d_b$, would affect the concentration difference $\delta$. While keeping the internuclear distance fixed at $d=200$ $\mu$m, we found that $\delta$ continuously increases with increasing $d_b$, as the outermost nuclei can attract more proteins from the boundary without having to compete with other nuclei (Fig.~\ref{fig3}(g)). However, when the distance to boundary $d_b$ becomes small enough, the concentration difference vanishes. In other words, the concentration at the end of S phase is maximal at the nucleus in the center of the domain.

\begin{figure*}[h!]
	\begin{center}
		\includegraphics[width=\textwidth]{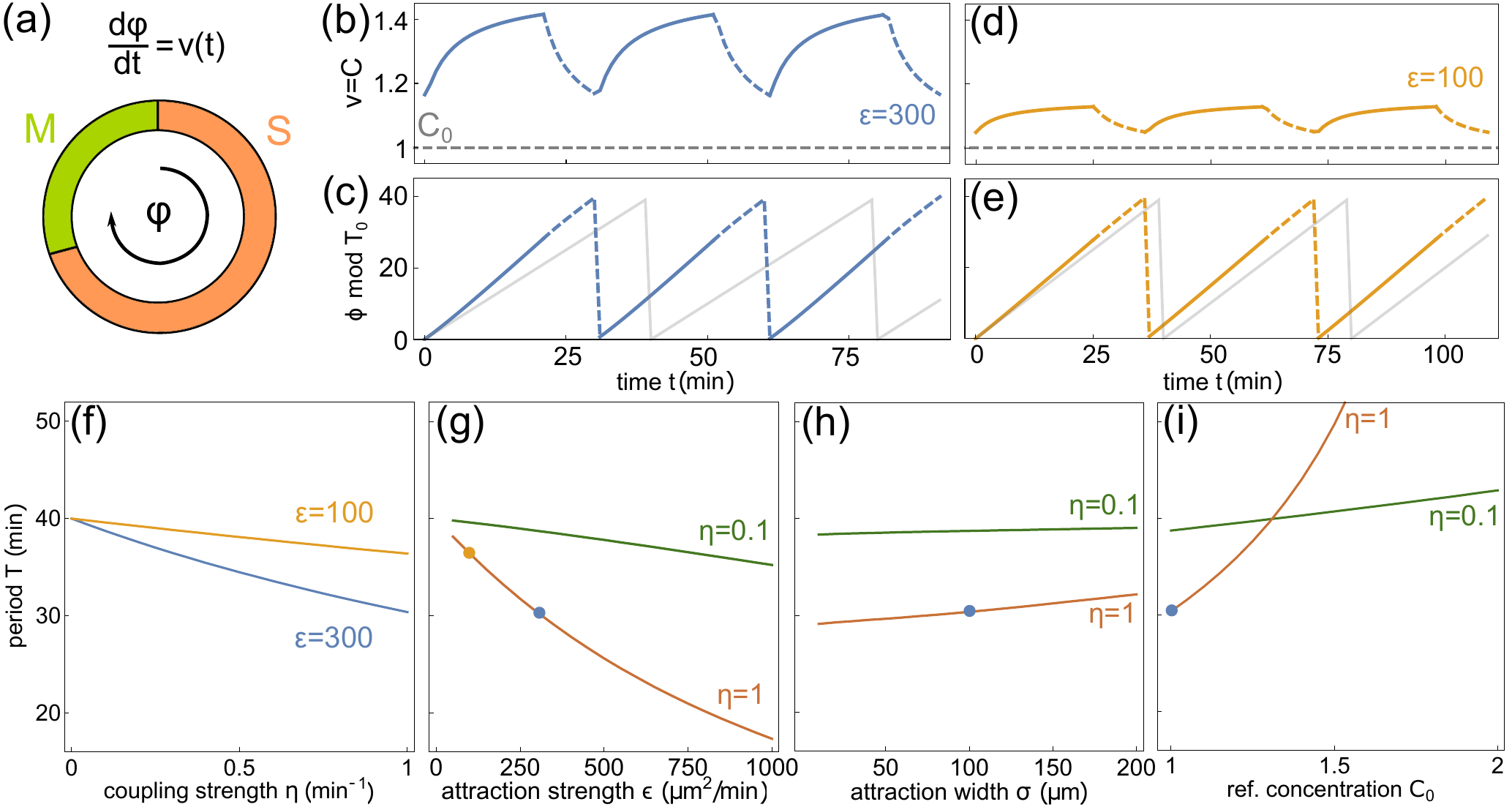}
	\end{center}
	\caption{
		The effects of nonzero coupling $\eta$ on the period of a single nucleus in a domain of size $L=1000$ $\mu$m. If not stated otherwise, parameters are chosen as in Table~\ref{tab1}. (a) Sketch of the cell cycle, consisting of S phase (fraction $\alpha$) and M phase with a total (reference) length $T_0$ of 40 minutes. The phase $\phi$ of a nucleus progresses through the cycle with a speed $v$, which is coupled to the local concentration via $\eta$ and $C_0$, determining the effective cell cycle period. (b) Concentration in the center as function of $t$ for attraction strength $\epsilon=300$ $\mu$m$^2$/min. The coupling strength is $\eta=1$ min$^{-1}$. Solid lines indicate that the nucleus is in S phase and dashed lines correspond to M phase. The gray dashed line denotes the reference concentration $C_0$. (c) The phase $\phi$ of the nucleus (mod $T_0$) corresponding to (b). The light gray line gives the phase for an uncoupled nucleus ($\eta=0$). (d) Similar as (b), for $\epsilon=100$ $\mu$m$^2$/min. (e) Similar as (c), for $\epsilon=100$ $\mu$m$^2$/min. (f) The effect of the coupling strength $\eta$ on the period, for both values of the attraction strength $\epsilon$ in (a). (g) The effect of attraction strength $\epsilon$ on the period, for coupling strength $\eta=0.1$ min$^{-1}$ (green) and $\eta=1$ min$^{-1}$ (brown).  The orange and blue dots correspond to an attraction strength of $\epsilon=100$ and $\epsilon=300$  $\mu$m$^2$/min, respectively. (h) Similar as (g), as function of the attraction width $\sigma$. (i) Similar as (g), as function of the reference concentration $C_0$. }
	\label{fig4}
\end{figure*}

\section{Model with a concentration-dependent cell cycle period}
The previous section illustrated that the competition between multiple nuclei to import proteins from a shared pool can lead to non-uniform spatial protein concentration profiles. In particular, the concentration was typically found to be higher towards to boundary of the domain. We then set out to study the system dynamics when the cell cycle period at the nuclear locations depends on the local protein concentration, i.e. $\eta>0$. Before tackling this situation of multiple nuclei, we again first look at the behavior of a single nucleus.

\subsection{Single nucleus}
As discussed in Section~\ref{sec:modeldef}, we associate a phase $\phi$ to the nucleus, which determines whether it is in S phase or M phase. This can be seen as a clock, where we define a speed $v$ determining how fast the cell cycle is traversed (see Fig.~\ref{fig4}(a)). Here, we let this speed $v$ change with the concentration according to (Eq.~\eqref{eq:vt}). This causes the effective period to change. For high concentrations $C$, the speed locally increases and thus the cell cycle period shortens. The situation without such coupling to the protein concentration, $\eta=0$, corresponds to $v(t)=1$, and thus $\phi=t$. Essentially, the clock phase corresponds to time, and the cell cycle period is constant (here $T=T_0=40$ min). Now let us fix the coupling strength at $\eta=1$ min$^{-1}$ and the reference concentration at $C_0=1$. In this case, the speed is simply $v(t)=C(\xi,t)$. Fig.~\ref{fig4}(b) shows the protein concentration $C$ (and thus the speed $v$) and Fig.~\ref{fig4}(c) the corresponding phase $\phi$ as function of time, for attraction strength $\epsilon=300$ $\mu$m$^2$/min. As the concentration at the nuclear positions increases with increasing import strength, the period is shortened as can be seen from the phase $\phi$ (Fig.~\ref{fig4}(d,e)). The case of constant period $T=T_0=40$ min is plotted in light gray for reference.

How much the cell cycle shortens depends on the coupling strength $\eta$ (Fig.~\ref{fig4}(f)). The decrease in the cell cycle period is larger for a larger attraction strength $\epsilon$, which is in correspondence with Fig.~\ref{fig4}(b-e). This effect is larger when the coupling strength $\eta$ is increased (see $\eta=0.1$ (green) vs. $\eta=1$ min$^{-1}$ (brown) in Fig.~\ref{fig4}(g)). When changing the nuclear attraction range $\sigma$, the effect is less pronounced: there is a small increase in cell cycle period for larger attraction width (Fig.~\ref{fig4}(h)). This small increase in period is due to the decrease in oscillation amplitude as shown in Fig.~\ref{fig2}.  Although the range of attraction is increased, the concentration at the end of S phase at the nucleus in the middle of the domain (which determines the period) is lower (see Fig.~\ref{fig2}(c) vs. Fig.~\ref{fig2}(a)). Finally, note that the concentration in Fig.~\ref{fig4}(b,d) is higher than the reference concentration $C_0$ in both phases of the cell cycle. This means that the cell cycle speeds up both in S phase and in M phase. By choosing a higher reference concentration in the range of the oscillations of $C$ (e.g. $C_0\approx1.3$ for $\epsilon=300$ $\mu$m$^2$/min), the cell cycle would speed up in certain parts of the cycle, and slow down in others. Indeed, the cell cycle speeds up when the local concentration $C(\xi)$ exceeds $C_0$, but slows down when $C(\xi)<C_0$. The overall cell cycle period thus increases with the reference concentration $C_0$, and this effect is stronger for larger coupling strength $\eta$ (see Fig.~\ref{fig4}(i)).

\begin{figure*}[h!]
	\begin{center}
		\includegraphics[width=\textwidth]{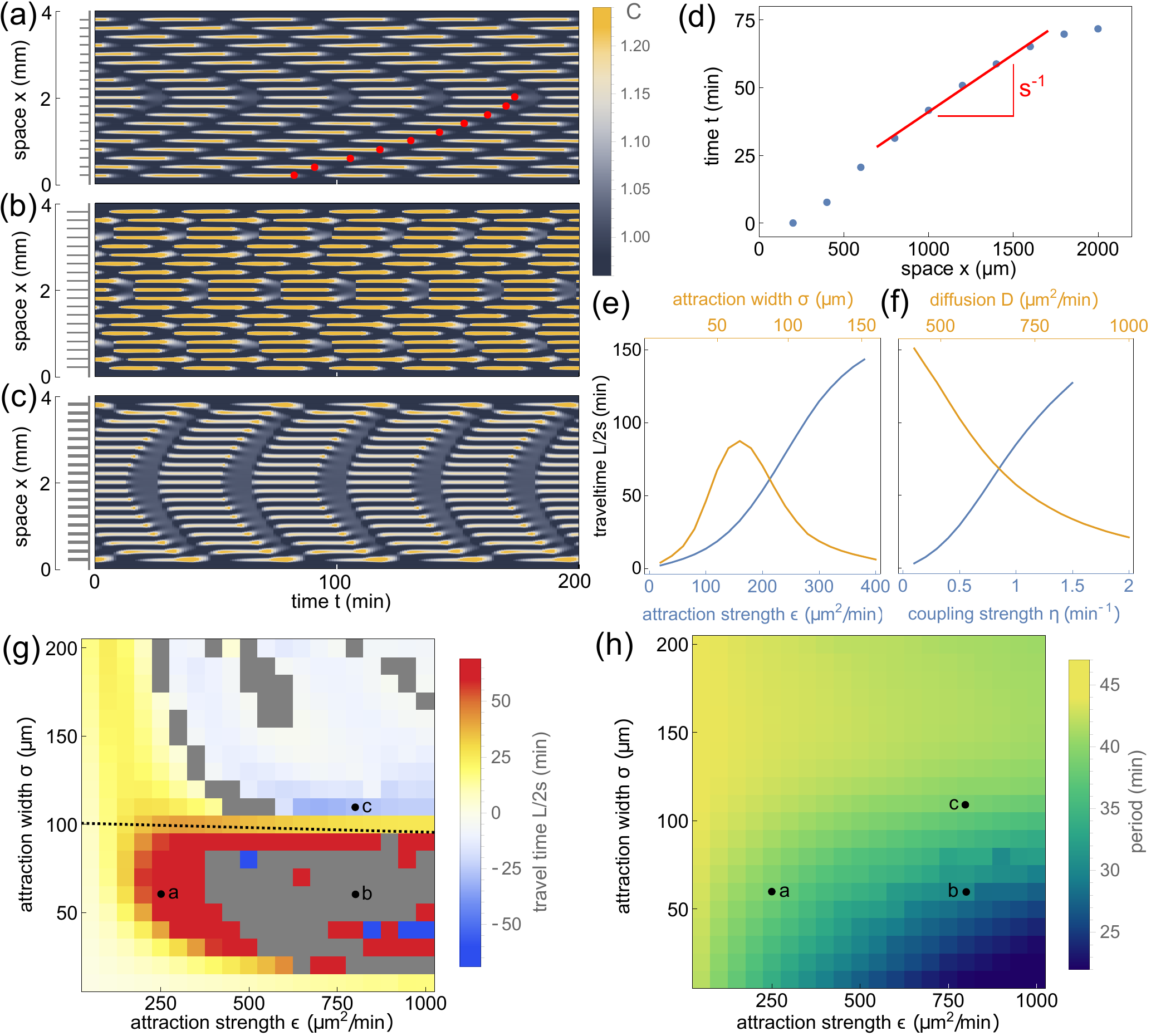}
	\end{center}
	\caption{
		(a) Concentration as function of space and time, after approximately 100 cycles for nuclear parameters $(\epsilon,\sigma)=(250,60)$. The red dots correspond to the entry into M phase. (b) Similar as (a) for $(\epsilon,\sigma)=(800,60)$. (c) Similar as (a) for $(\epsilon,\sigma)=(800,110)$. (d) Calculation of inverse wave speed $s^{-1}$, from the starting times of M phase at cycle 100 (time normalized to zero). The points correspond to the red dots in (a). A linear fit for nuclei 4--8 gives the inverse wave speed (and error on that fit) for this part of the domain. (e) Travel time of half the domain ($L/2s$) as function of nuclear parameters $\epsilon$ and $\sigma$. (f) Similar as (e), as function of coupling strength $\eta$ and diffusion coefficient $D$. (g) Travel time $L/{2s}$ as function of nuclear parameters $\epsilon$ and $\sigma$. Points a--c correspond to the three plots in (a--c). Gray areas correspond to points where the relative error is larger than 0.2. The dotted line corresponds to the line in Fig.~\ref{fig3}(c) between regions I and II. (h) Cell cycle period as function of nuclear parameters $\epsilon$ and $\sigma$. Points a--c correspond to the three plots in (a--c). }
	\label{fig5}
\end{figure*}

\subsection{Multiple nuclei}
When the cell cycle period is constant ($\eta=0$), multiple nuclei redistribute proteins in such a way that the local concentration at the end of S phase is higher at the boundary than in the center of the domain (see Figure~\ref{fig3}). Assuming that a higher local concentration will speed up the cell cycle ($\eta > 0$), one can expect the boundary to act as a pacemaker region spatially coordinating cell cycle timing. Indeed, Fig.~\ref{fig5}(a) shows that a mitotic wave (wave of mitotic entry) exists and originates at the boundary of the domain. The cell cycle periods at the nuclear locations, although initially different due to local concentration differences, have converged to the same value and the wave is able to synchronize the whole domain. Using the timing of mitotic entry of the different nuclei (red dots), the wave speed $s$ is estimated to be approx. 25 $\mu$m/min (see also Fig.~\ref{fig5}(d).). For a fixed cell cycle period ($\eta=0$), the concentration was always higher at nuclei located close to the boundary (see positive $\delta$ in Fig.~\ref{fig3}(c)), independent of the range and strength of attraction of the nuclei. However, for $\eta>0$ this is no longer the case. When increasing the attraction strength of the nuclei, clear wave behavior is lost (Fig.~\ref{fig5}(b)) as the nuclei still have different periods and fail to synchronize. When also increasing the attraction width of the nuclei, the wave direction can be reversed (Fig.~\ref{fig5}(c)). 

Fig.~\ref{fig5}(e--g) show in more detail how the different system parameters determine wave formation and the wave speed. Note that rather than plotting the the wave speed, we show the time it takes a wave to travel through half of the domain. The sign of this quantity $L/2s$ corresponds to the wave direction: when positive the wave originates from the boundary, while when negative it comes from the center of the domain. Fig.~\ref{fig5}(e) shows the travel time as function of the nuclear parameters $\epsilon$ and $\sigma$. Waves slow down as the attraction strength $\epsilon$ increases. The influence of the attraction width $\sigma$ is more complex. Waves are slowest at an intermediate $\sigma \approx 60 \mu m $, and speed up moving away from this value. For very small $\sigma$, the waves become near-synchronous as there is less interaction and for large $\sigma$ the potential functions overlap so much that there is less wave-like behavior. Fig.~\ref{fig5}(f) shows the travel time as function of the diffusion strength $D$ and coupling strength $\eta$. As is typical for traveling waves, the wave speed increases with the diffusion coefficient $D$. An increasing coupling strength between the local concentration $C$ and the cell cycle period has the inverse effect: waves are slower for larger values of $\eta$. For $\eta\to 0$ the travel time goes to zero as well (infinite wave speed), which indeed corresponds to the uncoupled situation where all nuclei go into S and M phase at the same time. A more detailed dependence of this travel time on the attraction strength $\epsilon$ and the attraction range $\sigma$ is shown in Fig.~\ref{fig5}(g). The reversal of the wave direction for sufficiently large values of the attraction strength and range is shown in blue. Whenever the error in computing a clear wave speed is too large, a gray color is assigned. The dotted line corresponds to the line in Fig.~\ref{fig3}(c) that separates regions I and II. In region I in the uncoupled situation ($\eta=0$), the maximum concentration is exactly located at the nucleus at the boundary, while in region II this position of maximum concentration moved more towards the center of the domain. In the latter case, the waves originate more often in the center when the coupling is turned on ($\eta>0$). Finally, in Fig.~\ref{fig5}(h), we show the period of the cell cycle in function of $\epsilon$ and $\sigma$. The period decreases with the attraction strength $\epsilon$, while it increases with the attraction range $\sigma$. This dependence is very similar to the case of a single nucleus in Fig.~\ref{fig4}. However, for larger attraction widths $\sigma$ (roughly the upper-half plane in Fig.~\ref{fig5}(g)) this dependence is less clear as the range of influence of the different nuclei largely overlap. 

\begin{figure*}[h!]
	\begin{center}
		\includegraphics[width=\textwidth]{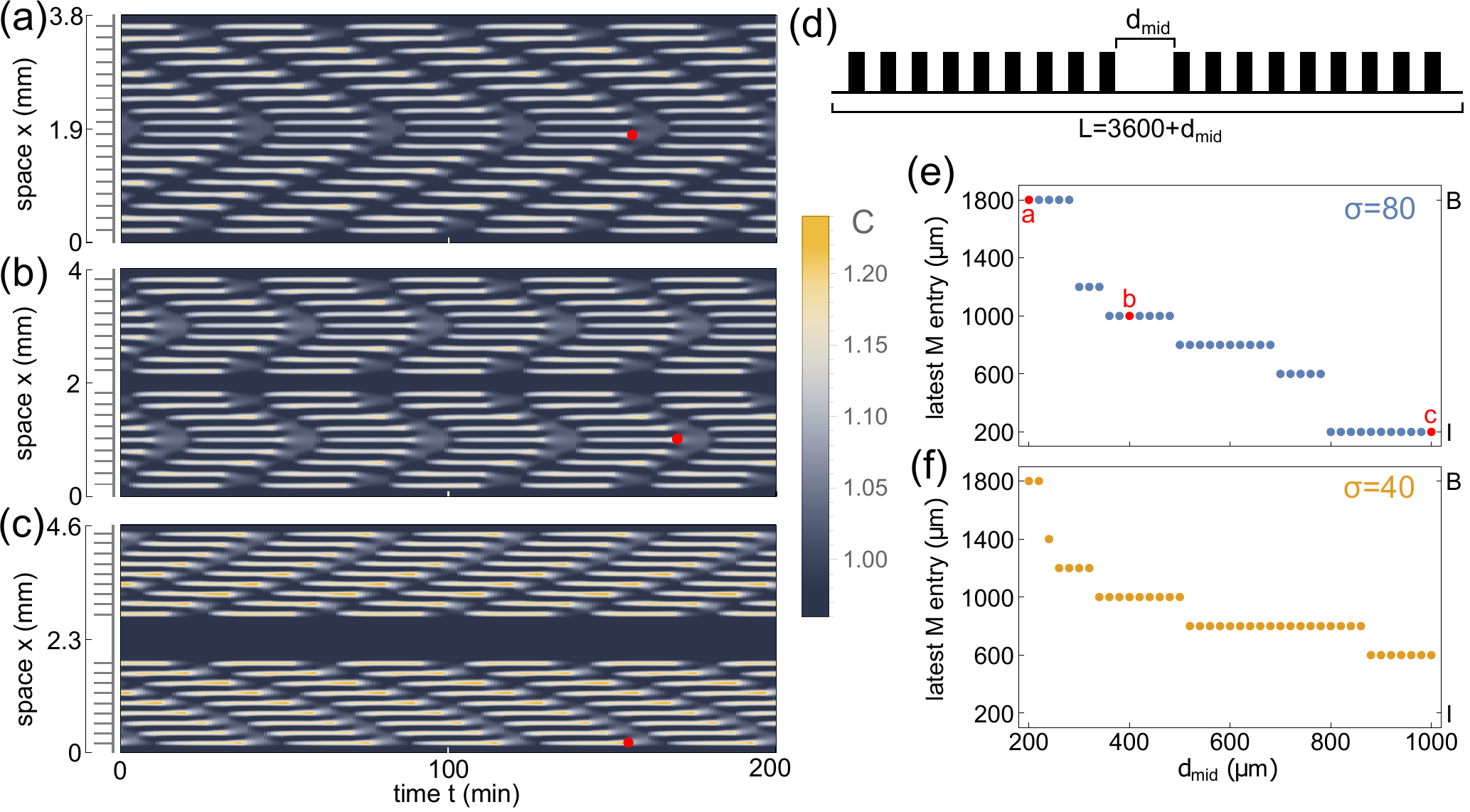}
	\end{center}
	\caption{
		The wave origin depends on nuclear positioning. (a) Concentration as function of space ($L=4000$ $\mu$m) and time (total of 200 minutes), after approximately 100 cycles. The nuclear parameters are fixed at $\epsilon=250$ $\mu$m$^2$/min and $\sigma=80$ $\mu$m, other parameters have standard values (Table~\ref{tab1}). Simulations for 18 nuclei with internuclear distance $d=200$ $\mu$m. The red dot is the nucleus with the latest M phase entry. (b) Similar as (a), for increased gap in the center of $d_{\text{mid}}=400$ $\mu$m. (c) Similar as (a), for increased gap in the center of $d_{\text{mid}}=1000$ $\mu$m. (d) Sketch of nuclear positioning with 18 equidistant nuclei except a changing distance $d_{\text{mid}}$. The total size $L$ of the domain changes with $d_{\text{mid}}$. (e) Location of the nucleus with the latest entry into M phase, as function of $d_{\text{mid}}$. The red dots correspond to the three situations in (a--c). The latest M entry at 200 $\mu$m (nucleus 1) corresponds to an internally driven wave (I) and a at 1800 $\mu$m (nucleus 9) to a boundary-driven wave (B). Simulations are done for $\sigma=80$ $\mu$m. (f) Similar as (e) for $\sigma=40$ $\mu$m.}
	\label{fig6}
\end{figure*}

The properties of mitotic waves (direction, speed) do not only depend on the nuclear parameters, but also on the positions of the nuclei. So far, we considered the effects of coupling on a system with equidistant nuclei (with an internuclear distance of 200 $\mu$m), which often gives rise to a mitotic wave originating at the boundary (Fig.~\ref{fig6}(a)). We wondered how the wave dynamics would be influenced by introducing a `gap' in the center of the domain by increasing the internuclear distance $d_{\text{mid}}$ between the two central nuclei. A sketch of this setup is shown in Fig.~\ref{fig6}(d). When the gap size $d_{\text{mid}}$ is increased, waves also originate from the nuclei close to this gap and they coexist with waves from the boundary (Fig.~\ref{fig6}(b)). For large enough gaps, a full reversal of the wave direction is observed, where the wave originating from the nuclei closest to the gap coordinate the whole domain (Fig.~\ref{fig6}(c)). These observations are in agreement with simulations in a similar model~\cite{Nolet2020}. Moreover, that study showed that small perturbations in the nuclear positioning do not affect the redistribution of proteins qualitatively. The red dots in Fig.~\ref{fig6}(a--c) correspond to the nucleus that has the latest entry into M phase in one of the waves. This location is directly correlated to the wave direction: when the middle nuclei (e.g. at 1800 $\mu$m in the first half of the domain) has the latest M entry, the wave comes from the boundary (Fig.~\ref{fig6}(a)). Vice versa, if the boundary nucleus, e.g. at 200 $\mu$m, has the latest M entry then the wave starts at the gap in the center (Fig.~\ref{fig6}(c)). For values in between, the two waves `compete' and the position of the nucleus with the latest M entry determines the wave that coordinates the largest part of the domain (Fig.~\ref{fig6}(b)). Figure~\ref{fig6}(e) shows which nucleus has the latest M entry, as function of the gap width $d_{\text{mid}}$. When decreasing the attraction width $\sigma$, the gap is no longer able to reverse the wave direction from a boundary-driven (B) wave to an internally driven (I) wave originating close to the gap (Figure~\ref{fig6}(f)). 

\section{Conclusions}
Early embryonic development requires a robust and quick progression through the cell cycle, in which cells replicate their DNA and divide into daughter cells. In some organisms, such as insects, fish and amphibians, these cells can be large, demanding spatial coordination to accomplish this. One example is the early embryo of the frog \textit{Xenopus laevis}, where it has been shown that mitotic waves are able to organize this process~\cite{Chang2013,Gelens2014}. Nuclei act as pacemakers of these waves~\cite{Afanzar2020,Nolet2020} and they could do this by locally concentrating cell cycle regulators~\cite{Nolet2020}. The role of the nucleus in the spatial redistribution of cell cycle regulators has been suggested before~\cite{Pines2010,Santos2012}. Also in the \textit{Drosophila} embryo, nuclei have been found to be important for the organization of cell cycle oscillations~\cite{HuangRaff1999,Deneke2019}. 

Here, we analyzed a generic computational model describing the redistribution of cell cycle regulators due to the presence of nuclei. The behavior is periodic: a nucleus imports proteins during S phase, whereas in M phase proteins can freely diffuse throughout the cytoplasm after the nuclear envelope has broken down. Furthermore, we assumed that the cell cycle oscillation frequency depends on the local concentration of an unspecified cell cycle regulator. We analyzed this model for a single nucleus and for multiple nuclei. Most cells, including those in the early embryo of \textit{Xenopus laevis}, only have one nucleus. However, in the \textit{in vitro} experiments with cell-free extract of the same organism, multiple nuclei self-organize in regular spatial patterns \cite{Nolet2020}. Moreover, in the early embryo of \textit{Drosophila melanogaster}, multiple nuclei share the same cytoplasm as well. Also in this \textit{in vivo} system mitotic waves have been observed, often originating at the boundary~\cite{FoeAlberts1983,Deneke2016}. Although we focus on the early development of these organisms, also in later stages cells can contain multiple nuclei. For example, in \textit{Drosophila} the positioning of multiple nuclei in muscle cells are studied experimentally and described with mathematical models~\cite{Manhart2018}.


When multiple nuclei are present in the same domain with a constant cell cycle period, our model shows that the redistribution of proteins leads to locally higher concentrations at nuclei close the boundary. The relative concentration difference between the maximum and the value in the center depends on the attraction strength $\epsilon$ and width $\sigma$. When nuclei are more effective in importing proteins (modeled with a higher $\epsilon$), the local concentration at the end of S phase increases. Although the attraction strength $\epsilon$ is changed for all nuclei at once, also the relative difference between the concentration at the boundary and in the center is increased. When increasing the attraction width $\sigma$ of the nuclei, the potential functions of the nuclei have a larger overlap and it becomes harder to distinguish between nuclei. The import by nuclei is weaker (since it is determined by the gradient of $V$), and the location of the concentration maximum starts to shift. For high enough values of $\sigma$, the maximum is no longer located at the outermost nuclei.

Similar as in \cite{Nolet2020}, we hypothesized that a concentration build-up in the uncoupled case leads to mitotic waves from the boundary when this concentration determines the cell cycle period. Indeed, we observe boundary-driven mitotic waves in our simulations of the coupled system. However, the properties of the wave -- its speed and direction -- highly depend on the attraction strength and width of the nuclei. For large enough attraction width $\sigma$, the wave direction can be reversed. Interestingly, for $(\epsilon,\sigma)$-values where the maximum is located at the outermost nuclei in the uncoupled case, waves tend to be always boundary-driven. Only when this maximum is shifted towards the center, the wave direction is reversed in the coupled situation. Moreover, we have shown that the wave direction does not only depend on the model parameters, it can also be affected by the positioning of the nuclei. When introducing a larger internuclear distance in the center, the wave direction can be reversed. For intermediate values there is coexistence of both waves. These findings are in agreement with those in \cite{Nolet2020}, in which it is shown that these types of waves are also observed in experiments. This work also showed the possibility of reversing the wave direction by increasing the (parameter equivalent to the) attraction strength $\epsilon$ for the nuclei in the center of the domain. However, in the model studied here we have not obtained simulations in which waves from the center were able to entrain the whole domain.

\begin{figure*}[h!]
	\begin{center}
		\includegraphics[width=\textwidth]{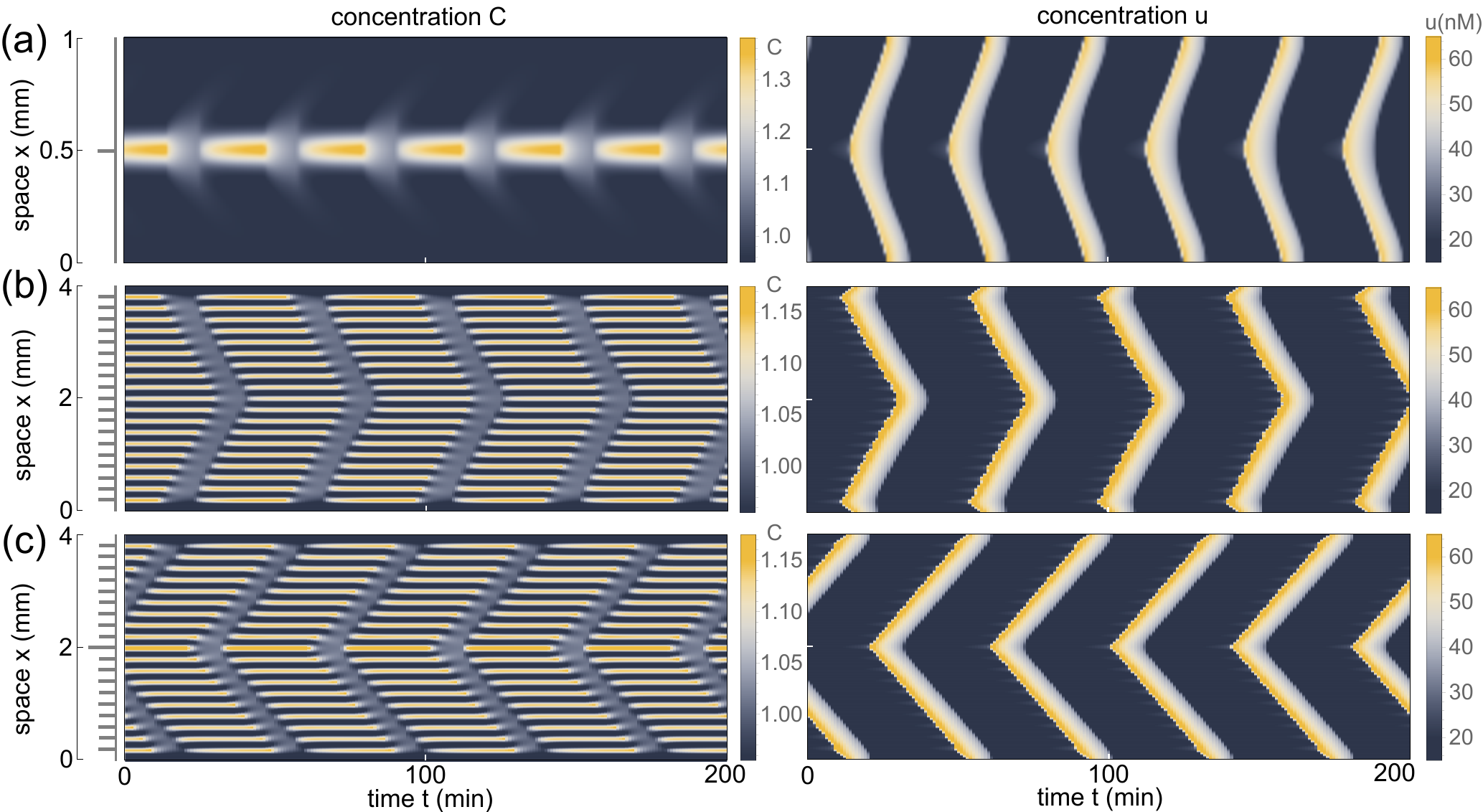}
	\end{center}
	\caption{
		Simulations with the import-diffusion model coupled to a cell cycle oscillator, shown after approximately 50 cycles. If not stated otherwise, parameter values are given in Table~\ref{tab1} and Table~\ref{tab2}. The nuclear parameters are fixed at $\epsilon=250$ $\mu$m$^2$/min and $\sigma=80$ $\mu$m and the reference concentration is $C_0=1.1$. (a) Concentration $C$ (left) and active Cdk1 ($u$, right) as function of space and time, for a single nucleus in a domain of length $L=1000$ $\mu$m.  (b) Concentration $C$ (left) and active Cdk1 ($u$, right) as function of space and time, for 19 nuclei in a domain of length $L=4000$ $\mu$m. (c) Similar as (b), where the middle nucleus has an attraction strength $\epsilon$ that is 20\% increased with respect to the other nuclei. }
	\label{fig7}
\end{figure*}

\section{Discussion}
\subsection{Waves in import-reaction-diffusion systems}
In this work we focused on the presence and properties of mitotic waves in a model where nuclei redistribute proteins and the local concentration determines the period of the cell cycle. This coupling is achieved by assigning a phase to each nucleus in the system and let the local concentration directly determine the phase speed. This is essentially a coupling between a periodic import-diffusion model to individual phase oscillators. Although this is sufficient to describe a lot of different types of (experimentally observed) behavior, cell cycle oscillations are usually not described by simple phase oscillators. 
There exist mathematical models that describe biochemical oscillations of Cdk1 activity in space and time with reaction-diffusion equations~\cite{Trunnell2011,Chang2013}. In combination with our equation for a cell cycle regulator $C$ this becomes a coupled system of the form
\begin{equation}
\left\{\begin{array}{ll}
\dfrac{\partial C(x,t)}{\partial t}&=D_C\nabla^2 C(x,t)+\nabla\cdot(C(x,t) \nabla V(u,x,t))\\[0.3cm]
\dfrac{\partial u(x,t)}{\partial t}&=D_u\nabla^2 u(x,t) + f(u,v,C)\\[0.3cm]
\dfrac{\partial v(x,t)}{\partial t}&=D_v\nabla^2 v(x,t) + g(u,v),
\end{array}\right.
\end{equation}
where $u$ denotes the concentration of active Cdk1 and $v$ the concentration of Cyclin B. The diffusion coefficients of the different variables are called $D_C$, $D_u$ and $D_v$, respectively. Note that $D_C$ corresponds to the coefficient $D$ as used before. The functions $f$ and $g$ are the reaction terms, describing the change of the concentration due to interactions with other proteins. The $u$-variable then defines whether the system is locally in S phase or M phase, via the potential function $V(u,x,t)$. Reversely, the concentration $C$ determines the speed of the cell cycle oscillation via the function $f$ (the reaction terms of Cdk1). This bidirectional coupling (turning the system into an import-reaction-diffusion system), replaces the phase oscillators where nuclei have a $C$-dependent phase speed. The details of this coupled model are described in Appendix~\ref{ap:CCOmodel}. This approach allows to define the phase (S/M) of the system everywhere in space, not only at the nuclear locations. 

Fig.~\ref{fig7} shows some initial results in this model. A single nucleus is now able to trigger waves in Cdk1 activity that propagate throughout the whole domain (Fig.~\ref{fig7}(a)). Although there is no wave-like behavior in the concentration $C$, it is this concentration that sets the pace of the oscillation in Cdk1, leading to mitotic waves through the domain. When simulating multiple nuclei, they all act as such pacemakers and the overall behavior is determined by differences in local concentration (Fig.~\ref{fig7}(b)). When the local concentration is higher at the boundary, these nuclei speed up the Cdk1 oscillations the most, and their waves eventually entrain the domain. Figure~\ref{fig7}(b) shows the concentration and Cdk1 activity for multiple nuclei in two cases: when all nuclei have the same attraction strength $\epsilon$ and Fig.~\ref{fig7}(c) when the middle nucleus has a 20\% higher attraction strength. In the latter case, the wave direction reverses, as the local concentration in the center is increased, leading to a higher cell cycle frequency. These observations are in agreement with an earlier theoretical analysis of competing pacemakers~\cite{Nolet2020chaos}.

\subsection{Future work}
One way to extend this work would be to further study the import-reaction-diffusion system proposed in the previous section. Whereas we focused on a specific example, different reaction parts of the model can be chosen to investigate the effect on the resulting oscillation and wave dynamics. In our example we used a cell cycle oscillator of the relaxation-type, built on underlying bistability. It would be interesting to see how different types of oscillators (i.e. sinusoidal vs. relaxation oscillators) affect the wave properties, which has recently been characterized in the context of pacemaker-driven waves in reaction-diffusion systems~\cite{Rombouts2020}. The type of coupling used in the model offers another possibility to study in more detail. We currently let certain parameters of the reaction part depend on the concentration $C$, but introducing the coupling elsewhere could also influence the observed dynamics of the system. 

Although this work focuses on simulations in one spatial dimension, the general setup of the model allows for studies in two or three dimensions as well. Quasi-2d experiments with \textit{Xenopus} cell-free extract recently showed similar mitotic wave behavior~\cite{Afanzar2020,Nolet2020} as in quasi-1d experiments~\cite{Nolet2020}. This model could be used to describe some of the observed phenomena there. In two dimensions, the waves generated by pacemakers can lead to target patterns. Moreover, in 2d, there is much more freedom to explore different sizes and geometries of pacemakers (e.g. nuclei), which in 1d has shown to be of importance for the dynamics of the system~\cite{Nolet2020chaos}. 

In our model, the positions $\xi_i$ of the nuclei can be chosen freely, but they do not change in time. It would be interesting to include the motion of nuclei in the cytoplasm, e.g. by forces due to the presence of microtubules. This could be done by explicitly including microtubules into the model. However, this would significantly increase computational time. As a first step, the effects of microtubules can be incorporated indirectly, by adding a force between nuclei and between nuclei and the boundary. This force, depending on the distance, describes the average effect of microtubules and when exerted on a nucleus it will move in space accordingly. This approach would be similar to the study of nuclear positioning in muscle cells, where multiple nuclei share the same cytoplasm as well~\cite{Manhart2018}.  

\section*{Acknowledgements}
We thank Jan Rombouts, Arno Vanderbeke and Daniel Ruiz Reyn\'es for valuable feedback on the manuscript. This work was supported by the Research Foundation - Flanders (FWO, grant GOA5317N) and the KU Leuven Research Fund (C14/18/084).

\appendix
\section{Appendix: Additional information about the numerical methods}
\label{ap:numerics}
\subsection{Boundary conditions}
In section \ref{sec:modeldef}, we defined the model and discussed the boundary condition in general. Here, we discuss the implementation of this and the numerical details of solving the model equation. In one dimension, we can write the equation for $C$ as
\begin{equation}
\frac{\partial C}{\partial t}=D\frac{\partial^2 C}{\partial x^2}+\frac{\partial C}{\partial x}\frac{\partial V}{\partial x}+C\frac{\partial^2 V}{\partial x^2},
\end{equation}
or equivalently in flux-form as
\begin{equation}
\frac{\partial C}{\partial t}+\frac{\partial J}{\partial x}=0,
\end{equation}
with the flux $J$ given by
\begin{equation}
J=-\left(D\frac{\partial C}{\partial x}+C\frac{\partial V}{\partial x}\right).
\end{equation}
We write the functions $C,V$ as $C=C(x,t)$ and $V=V(x,t)$ and solve the model equation numerically on a grid for $x,t$ with distances $\Delta x$ and $\Delta t$. The derivative of $C$ with respect to $t$ at $t=t_0$ is numerically approximated by
\begin{equation}
\frac{\partial C}{\partial t}\Big|_{x=x_i,t=t_0}\approx\frac{C(x_i,t_0+\Delta t)-C(x_i,t_0)}{\Delta t}
\end{equation}
for all grid points $x_i$, which is equivalent to an Euler-forward method. This derivative should be equal to $-\frac{\partial J}{\partial x}$, which is calculated via
\begin{equation}
\frac{\partial J}{\partial x}\Big|_{x=x_i}\approx-\frac{J(x_{i+1/2},t)-J(x_{i-1/2},t)}{\Delta x}
\end{equation}
for all $t$. The quantity $J(x_{i+1/2},t)$ denotes the flux of $C$ from $x_i$ to $x_{i+1}$ at time $t$. This flux is calculated via 
\begin{equation}
\begin{array}{l}
J(x_{i+1/2},t)=D\dfrac{C(x_{i+1},t)-C(x_{x_i},t)}{\Delta x}\\[0.3cm]
\qquad+\dfrac{(C(x_i,t)+C(x_{i+1},t))(V(x_{i+1},t)-V(x_i,t))}{2\Delta x}.
\end{array}
\end{equation}
Using the equation with the flux for solving the model equation has two main advantages: (i) the quantity $C$ is conserved by construction and (ii) a zero-flux boundary condition can be easily implemented. The latter can simply be done by defining the fluxes $J(x_{-1/2},t)$ and $J(x_{N+1/2},t)$ to be zero, i.e. there is no flux through the boundary at the points $x_0=0$ and $x_N=L$, the edges of the domain.

\subsection{Simulation time}
Space and time were calculated in micrometers and minutes, respectively. Normally, grid sizes of $\Delta x=5$ $\mu$m and $\Delta t=0.01$ min were taken. These values were chosen large enough to avoid too long computational time, while at the same time small enough to avoid a significant increase in numerical errors. In simulations with a constant cell cycle period, i.e. $\eta=0$, the system reaches a limit cycle in only a few cycles. Results are plotted when this limit cycle was reached. For the coupled system $\eta>0$, the system is most cases constantly in transient. For some parameter values the system reaches a limit cycle, however this was often not the case. The waves that we observe in the coupled situation, are analyzed after a simulation of 100 cycles. This is long enough for mitotic waves to form, and the properties are stable on small time scales (e.g. similar values would be obtained at cycle 90 or 110). However, on longer timescales there are still changes in behavior. When simulating much longer (in the order of 100s to 1000s of cycles), the speed and even the direction might become different. The system is constantly changing: even after more than 1000 cycles it often has not reached a limit cycle. Interestingly, in the proposed import-reaction-diffusion system in the discussion, the system does reach a limit cycle. This again hints at the importance of the oscillator type. 

\section{Appendix: Coupling the import-diffusion model to a biochemical oscillator}
\label{ap:CCOmodel}
Cell cycle oscillations can be described in space and time by two partial differential equations for the concentrations of Cdk1 and Cyclin B. When setting $u=[\text{Cdk1}]$ and $v=[\text{Cyclin B}]$, the model is defined by
\begin{equation}
\label{eq:CCO}
\left\{\begin{array}{lll}
\dfrac{\partial u}{\partial t} &= D_u\nabla^2 u 
+ \left(a_1+b_1\frac{u^{n_1}}{E_1^{n_1}+u^{n_1}}\right)(v-u) \\[0.3cm]
&-\left(a_2+b_2\frac{E_2^{n_2}}{E_2^{n_2}+u^{n_2}}\right)u \\[0.3cm]
&-\left(a_3+b_3\frac{u^{n_3}}{E_3^{n_3}+u^{n_3}}\right)u+k\\[0.5cm]
\dfrac{\partial v}{\partial t} &= D_v\nabla^2 v 
-\left(a_3+b_3\frac{u^{n_3}}{E_3^{n_3}+u^{n_3}}\right)v +k.
\end{array}\right.
\end{equation}
In these equations, consisting of a diffusion term and one or more reaction terms, $k$ denotes the synthesis rate and the diffusion coefficients are $D_u$ and $D_v$ for the respective variables. The other parameters in the model correspond to interactions with other proteins. These parameters are divided in three classes (numbered 1,2,3), which are present in different terms in the equations (which we can give the same number). Term 1 then corresponds to the interaction with the kinase Wee1, term 2 to the phosphatase Cdc25 and term 3 (present in both equations) corresponds to degradation via APC/C. Standard values of the parameters involved are given in Table~\ref{tab2}. For the standard values, the $u$-term (i.e. Cdk1) toggles between a state with low activity (S phase) and high activity (M phase). For more details on the model equations and the type of oscillations resulting from those, we refer to \cite{Chang2013,Nolet2020} and references therein. 

\begin{table}[b!]
	\begin{center}
		\begin{tabular}{c l l || c l l }
			\toprule
			Par.   & Value	 & Unit   &Par.  & Value	 & Unit   \\
			\midrule
			$a_1$    & 0.8 	& min$^{-1}$ 	&$b_1$   & 4 	& min$^{-1}$  \\
			$a_2$    & 0.4	& min$^{-1}$ 	&$b_2$   & 2 	& min$^{-1}$  \\
			$a_3$    & 0.01 & min$^{-1}$  	&$b_3$   & 0.06 & min$^{-1}$  \\
			$E_1$    & 35 	& nM  			&$n_1$   & 11 	&   \\
			$E_2$    & 30 	& nM  			&$n_2$   & 3.5 	&   \\
			$E_3$    & 32 	& nM  			&$n_3$   & 17 	&   \\
			$D_u$    & 600 	& $\mu$m$^2$/min  &$k$    	 & 1.5	& nM/min  \\
			$D_v$    & 600 	& $\mu$m$^2$/min  &    &  	&   \\
			\bottomrule
		\end{tabular}
	\end{center}
	\caption{
		Standard parameter values and their units for the cell cycle oscillator model.}
	\label{tab2}
\end{table}

For the parameter values in Table~\ref{tab2}, the solution for $u$ switches between a state of low concentrations, with values in the range of 10--25 nM, and a state of high concentrations, in the range of 40--65 nM. Instead of having a reference period ($T=40$ min) with a phase oscillator for all nuclei, we let the $u$-concentration determine whether the system is in S phase or M phase. Therefore, we set a threshold at $u=30$ nM, with lower values corresponding to S phase and higher values to M phase. In other words, the phase is not only defined for the nuclei, but everywhere in space. Whether a nucleus is importing is then dependent on the local value of $u$, via the function $F_i(t)$ for all different nuclei. Recall (from Section~\ref{sec:modeldef}) that the potential function $V_i$ of a nucleus could be written as $V_i(x,t)=F_i(t)G_i(x)$, where $G_i$ is an exponential function in space determined by the attraction strength and width. This will be kept the same, however the time-dependent function $F_i$ is now defined as
\begin{equation}
F_i(t)=\left\{
\begin{array}{ll}
1&\text{if } u(\xi_i,t)<30\text{ nM}\\
0&\text{if } u(\xi_i,t)\geq 30\text{ nM},
\end{array}\right.
\end{equation} 
thus the phase is dependent on the concentration of active Cdk1 at the position of the nucleus. The coupling of the local concentration to the local oscillation frequency is obtained via the parameters $a_1$ and $b_1$, belonging to the interaction with Wee1. In the original model, this coupling was achieved by letting the local concentration at a nucleus determine the speed $v_i$ of the phase oscillator. In the model for cell cycle oscillations, it is known that an increase in Wee1 via leads to faster oscillations~\cite{Nolet2020}. This can be achieved by multiplying the corresponding term,
\begin{equation}
a_1+b_1\frac{u^{n_1}}{E_1^{n_1}+u^{n_1}},
\end{equation}
with a certain factor $\beta_i$, depending on the local concentration $C(\xi_i,t)$. This is equivalent to scaling $a_1$ and $b_1$ with the same factor. This factor $\beta_i$ is then given by
\begin{equation}
\beta_i(t)=1+\eta(C(\xi_i,t)-C_0),
\end{equation}
which is a similar equation as for the speed of the phase oscillator before. Again, $\eta$ is a measure for the coupling strength and $C_0$ is a reference concentration.

\bibliographystyle{apsrev}


\end{document}